\documentclass[secnumarabic,amssymb,nobibnotes,aps,prd,showpacs]{revtex4}%
\usepackage{graphicx}
\usepackage{dcolumn}
\usepackage{bm}
\usepackage{amsmath}%
\usepackage{amsfonts}%
\usepackage{amssymb}

\begin{document}

\title{Is there a GeV counterpart in the Fermi GBM events? }

\author{C. R. A. Augusto, C. E. Navia, K. H. Tsui, and H. Shigueoka}
\affiliation{Instituto de F\'{\i}sica, Universidade Federal Fluminense, 24210-346,
Niter\'{o}i, RJ, Brazil} 

\date{\today}
\begin{abstract}
In this paper we report a possible GeV counterpart observed at ground level as muons (or "photo-muons") of two Fermi gamma burst monitor (GBM) events, trigger bn081120618 and trigger bn081124060. In both cases, the trigger coordinates are within the field of view of the vertical Tupi telescope located at sea level and inside the South Atlantic Anomaly (SAA) region. We show that despite the first trigger being classified as GRB (due to a gamma ray burst), with an 100\% probability assigned by GBM Flight Software, the trigger time 14:49:31 UT happened during particle precipitations in the SAA, which in most cases is $\sim 12h$ UT to $\sim 22h$ UT, and the probability of the trigger being attributed to particle precipitation, according to our analysis is 25\%. The second Fermi trigger is also classified as GRB, with a 100\% probability assigned by GBM Flight Software. The trigger time is 01:26:07 UT, which is outside the schedule of precipitation of particles. Another important observed characteristic is the existence of a ground level enhancement (GLE) with a sharp peak ($5.1\sigma$) coinciding with the Fermi trigger within the 10-second counting interval (raw data) of the vertical Tupi telescope. In addition it is possible to identify other GLEs before and after the trigger occurrence. In both cases, the scenario is similar to the long-duration GeV GRBs observed by EGRET within the BATSE field of view.
\end{abstract}


\maketitle

\section{Introduction}
The Tupi experiment comprises two small telescopes, one with vertical orientation and the other with $45^0$ in relation to the vertical and pointed to the West. They are mounted on bases of plastic scintillators and detect muons ($E_{\mu}>1GeV$) at sea level produced by cosmic rays (mostly protons) including gamma-rays with energies above the pion production ($\sim 10 GeV$) in the Earth's atmosphere. The Tupi experiment is located in Niteroi City, Rio de Janeiro, Brazil (22S, 43W), inside the so-called South Atlantic Anomaly (SAA) region. Each telescope has a field of 0.27 sr and a duty cycle greater than 95

On the other hand, according to Measurements of Pollution in the Troposphere (MOPITT))\cite{mopitt}, the SAA is a very low magnetic field region at ground. In the IGRF95 data \cite{igrf95}, the magnetic field strength in the SAA region (26S, 53W) is 24000 nT, around two times lower than the transversal magnetic fields in the Antarctic High region (70S, 140E) which is 60000 nT. In addition, the Earth's magnetic field deflects the charged particles of the shower initiated by a gamma ray. This deflection is caused by the component of the Earth's magnetic field perpendicular to the particle trajectory. This effect results in a decrease in the number of collected particles and therefore in the telescope's sensitivity.

The radius of curvature, $R$, for instance of a vertical positive muon traveling downward in the atmosphere with momentum $p$, because of the Earth's transverse magnetic field, $B_T$, is $R=p/(eB_T)$. As the muon travels it will be shifted a (horizontal) quantity $\delta x$ in the direction perpendicular to $B_T$. A relation for $\delta x$ in the first order in $z/R$, where $z$ is the height of the atmosphere where the muon is generated, can be obtained as $\delta x \sim z^2/R= z^2ceB_T/p$. Thus the muons are shifted by a quantity that depends on the travel distance $z$ but also on the momentum $p$ of the muon. The $\delta x$ in the SAA area is at least 50\% smaller than the $\delta x$ outside the SAA area. This means that the sensitivity of particle telescopes is highest in the SAA region, because in this region the transverse magnetic field is very small, and even smaller than the average value of the polar regions.  
This characteristic offers the Tupi muon telescopes located in the SAA region a low rigidity of response to primary and secondary charged particles ($\geq 0.4$ GV), and it offers the opportunity to observe transient events of small scale \cite{navia05,augusto05}.

The major background interference in gamma ray detection by spacecraft detectors is cosmic charged particles, mainly protons, electrons, and a smaller number of neutrals and Earth albedo photons. Thus the space gamma ray detectors are inside an Anti-Coincidence Detector (ACD), a device which identifies and vetoes charged particles. Spacecrafts such as Fermi \cite{gargano08} orbit Earth at altitudes of about 550 km with latitudes between $\pm 26^0$, and will encounter the SAA region. When passing through the SAA, Fermi can be subjected to fluxes up to 700\% higher than average.
The open magnetosphere in the SAA region also propitiates the precipitation the particles with energies above the pion production threshold, because they produce muons in the Earth's atmosphere. They look like sharp peaks in the muon counting rate.

On some occasions we have observed that the muon peaks produced by high energy particle precipitations coincide with the occurrence of Swift-BAT, MILAGRO \cite{augusto08a}, and Fermi GBM triggers \cite{augusto08b}. Swift and Fermi are two space-borne GRB detectors and MILAGRO was a ground-based GRB detector. In all cases the trigger coordinates are close to or inside the field of view of the telescopes. These triggers are probably produced by precipitation of charged particles in the SAA region and not due to gamma ray bursts.

The precipitation of high energy particles in the SAA region is subject to fluctuations \cite{augusto08a}. For instance, fluctuations exist over several days when no precipitation of particles is observed in this region. The origin of these fluctuations can be described using the "Open Magnetosphere" Model \citep{dungey61,debrito05}, based on a "reconnection process" that takes place at the front (day-side) of the magnetosphere. 
The precipitation of high energy particles in most cases begins three hours after sunrise and finishes one hour after sunset \cite{augusto08a}. However this behavior is subject to seasonal variations.
\begin{figure}[th]
\vspace*{-1.0cm}
\hspace*{-0.0cm}
\includegraphics[clip,width=0.8
\textwidth,height=0.7\textheight,angle=0.] {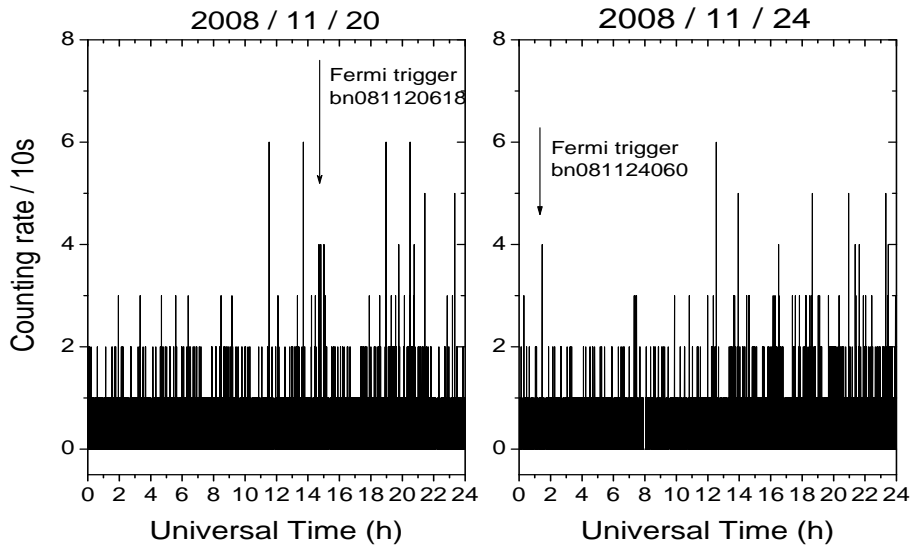}
\vspace*{-8.0cm}
\caption{Time profile rates were observed in the vertical Tupi Telescope on $2008/11/20$ (left) and $2008/11/24$ (right). In most cases, the sharp peaks in the muon counting rate after $\sim 12h$ UT are due to high energy ($>0.1$ GeV) particle precipitations in the SAA region. The vertical arrows represent the Fermi GMB trigger occurrences.}
\end{figure}

In this paper, we report a search for the GeV counterpart of two Fermi GBM events, trigger bn081120618 and trigger bn081124060 \cite{fermiCat}, observed at ground level by the vertical Tupi muon telescope. The main characteristic observed in both cases is the occurrence of several muon enhancements (sharp peaks) all with high confidence levels in the muon counting rate, around both Fermi triggers. Fig. 1 summarizes the situation, showing the vertical telescope output for two days: November 20, 2008 and November 24, 2008, in the left and right panels respectively.
They represent the muon counting rate at every 10 seconds (raw data); in both cases, the vertical arrows represent the Fermi GBM trigger occurrence.
 
As happened in the two previous events reported in \cite{augusto08c}, an important characteristic observed in these two Fermi events is that their trigger coordinates are inside the effective field of view of the vertical muon telescope.  So far, we have three Fermi events whose GBM triggers coordinates are inside the field of view of the Tupi telescope,
the first was reported in \cite{augusto08c} and the last two are detailed below.

\section{Fermi trigger bn081120618}

Fermi event trigger bn081120618 is the second whose trigger GBM coordinates are within the field of view of the vertical telescope. Fig. 2 shows the telescope axis equatorial coordinates together with the Fermi trigger coordinates. The "ellipses" represent the effective field of view of the vertical and inclined telescopes and the open circle represents the Sun's position. We can see that the direction of the axis of the inclined telescope, practically points in the direction of the Sun.

\begin{figure}[th]
\vspace*{-1.0cm}
\hspace*{-2.0cm}
\includegraphics[clip,width=0.5
\textwidth,height=0.5\textheight,angle=0.] {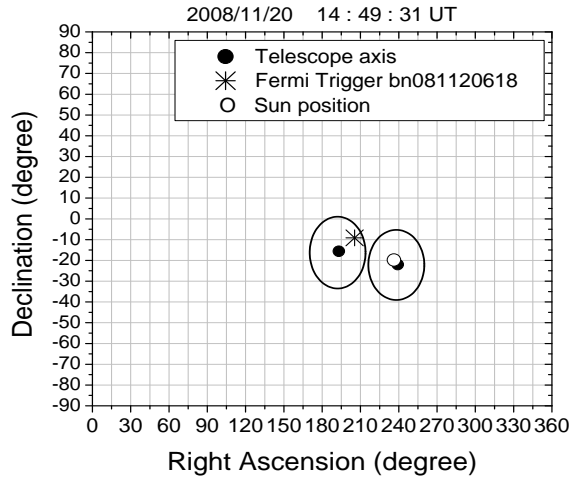}
\vspace*{-4.5cm}
\caption{Equatorial coordinates showing the position of the vertical and inclined telescope axis.. The "ellipses" represent the field of view of the telescopes, the asterisk is the position (coordinates) of Fermi GBM trigger bn081120618, and the open circle represents the Sun's position.}
\end{figure}

\begin{figure}[th]
\vspace*{-0.0cm}
\hspace*{-2.0cm}
\includegraphics[clip,width=0.5
\textwidth,height=0.5\textheight,angle=0.] {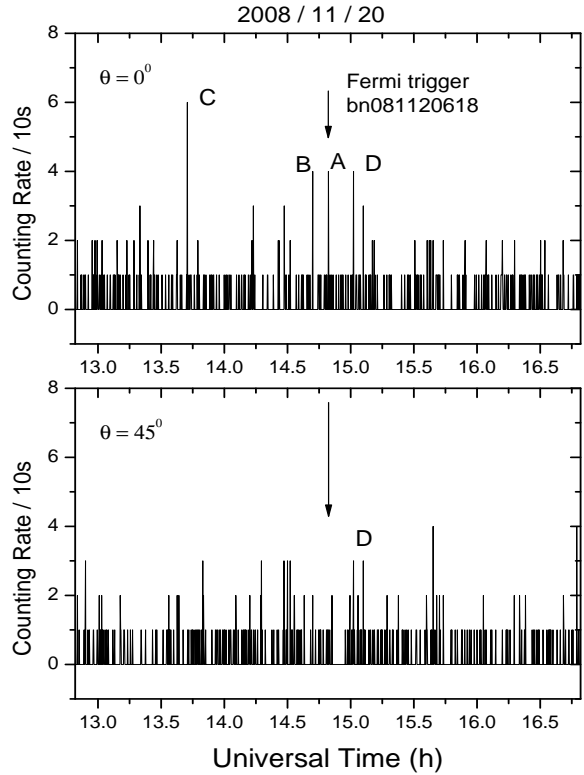}
\vspace*{-1.0cm}
\caption{Time profile rates at every 10s: (upper panel) for the vertical telescope and (lower panel) for the inclined telescope, on November 20, 2008, around 2 h before and 2 h after the Fermi GBM trigger occurrence, indicated by a vertical arrow.}
\end{figure}

According to the Fermi GBM Trigger Catalog (version 0), the probability of event trigger bn081120618 being due to a solar flare was 0.8588. However, we examine a possible observation at ground of its GeV counterpart, whose signature is difficult to interpret as being originated by a solar flare. The occurrence of the Fermi trigger is during a plentiful particle precipitation in the SAA, on November 20, 2008, as is shown in Fig. 1, where sharp peaks with high confidence levels can be observed. The trigger time is 14:49:34 UT and it is inside the schedule ($\sim 12h\;UT$ to $\sim 22h\;UT$), where precipitation of particles is observed commonly in the SAA region \cite{augusto08a}. Probably, the fact that the Fermi trigger happens in the daytime and with the positioned Sun close to the trigger coordinates (see Fig. 2) has contributed to the trigger being classified as SFL, that is, with a high probability of being originated by a solar flare. In addition, according to the SoHO CELIAS/SEM X-ray sensor \cite{soho}, apparently the Sun did not undergo an X-ray flaring activity at the trigger time, or close to it. However, in the Fermi GMB Trigger Catalog (version 3) the trigger was reclassified to the category of GRB with a reliability of 100\%. In order to see better the fine structure of this possible assocoation, the time profile rate in both (vertical and inclined) telescopes around two hours before and two hours after the trigger occurrence is shown in Fig. 3. The omni directianality behavior of the particle precipitation in the SAA can be useful to test the origin of the muon peaks around the Fermi trigger, if they are produced by precipitations, they must appear in both telescopes.

Following the muon counting rate in the vertical telescope (upper panel of Fig. 3) is possible to see a peak (peak A, $7.5\sigma$) coinciding with the Fermi trigger within the 10-second counting interval (raw data). There are also at least three  peaks: at 7.8 minutes (peak B, $7.5\sigma$) and at 65.4 minutes (peak C, $11.5\sigma)$ before the Fermi trigger, and a peak at 12.0 minutes (peak D, $7.5\sigma$) after the Fermi trigger. Of these peaks only the peak D appears (in coincidence) in the inclined detector and with a lower confidence level. This means that the other peaks have chance of they be produced by a GeV counterpart of a keV GRB and whose trigger is bn081120618. The scenario is similar to the long-duration GeV GRBs observed by EGRET within the BATSE field of view \cite{hurley94}.
Even so, the trigger is in the schedule of the particle precipitation in the SAA, 
and of the analyzes above described, we considered for this trigger a probability of 75\% of being produced by GRB.

\section{the Fermi trigger bn081124060}

Fermi trigger bn081124060 has been cataloged as GRB (due to a gamma ray burst) with a probability of 93.73\% by the GBM flight software (version 0) and with a probability of 100\% in the version 1, and it is the third whose trigger coordinates are within the field of view of the vertical telescope. Fig. 4 summarizes the situation, showing the telescope axis equatorial coordinates together with the Fermi trigger coordinates.
\begin{figure}[th]
\vspace*{-1.0cm}
\hspace*{-2.0cm}
\includegraphics[clip,width=0.5
\textwidth,height=0.5\textheight,angle=0.] {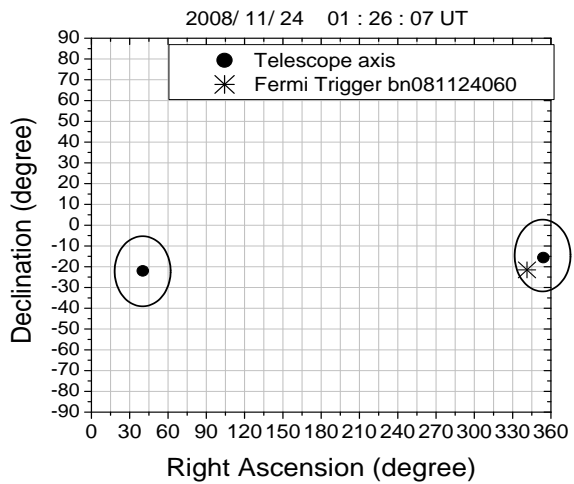}
\vspace*{-4.5cm}
\caption{Equatorial coordinates showing the position of the vertical and inclined telescopes. The "ellipses" represent the field of view of the telescopes, the asterisk is the position (coordinates) of the Fermi GBM trigger bn081124060.}
\end{figure}

As shown in Fig. 1 (right panel), on November 24, 2008, the Tupi telescopes also registered particle precipitation. As expected, the precipitation is confined in its habitual schedule with a beginning at $\sim 12h$, and the trigger time at 01:26:07 UT is outside this region. However, in order to see the details of this peculiar event, we also show in Fig. 5 the muon counting rate in the vertical telescope around 1.5 h before and 1.5 h after the Fermi trigger, for two counting rates: 10 s (upper panel) and 1 m (lower panel) respectively.

\begin{figure}[th]
\vspace*{-1.0cm}
\hspace*{-2.0cm}
\includegraphics[clip,width=0.5
\textwidth,height=0.5\textheight,angle=0.] {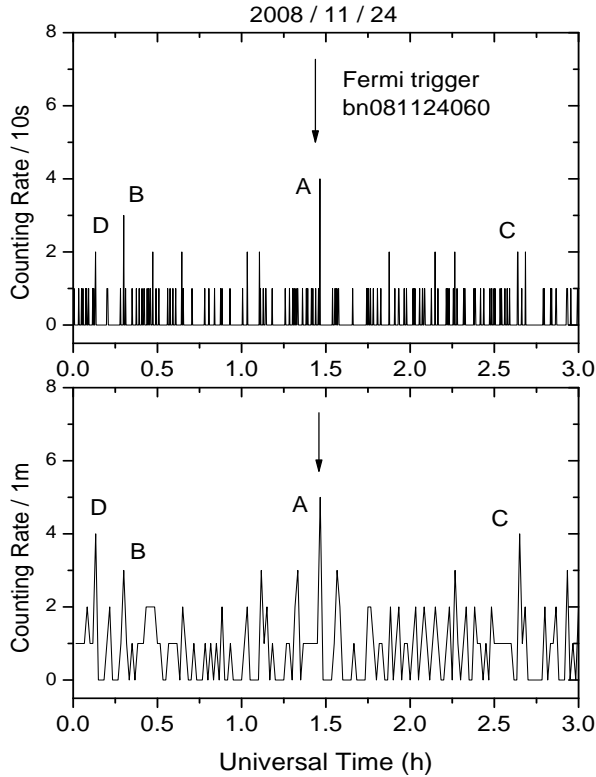}
\vspace*{-1.0cm}
\caption{Time profile rates at every 10 s (upper panel) and every minute (lower panel), both in the vertical Tupi telescope on November 24, 2008, around 1.5 h before and 1.5 h after the Fermi GBM trigger occurrence, indicated by a vertical arrow.}
\end{figure}

From this figure the existence of a sharp peak (peak A) is evident with a confidence level of $5.1\sigma$ and coinciding with the Fermi trigger within the 10-second counting interval (raw data) of the vertical Tupi telescope. 
In the counting rates at every minute, peak A is enhanced to $6.2\sigma$, which means that its duration is around $\sim 1m$. In addition, there are at least two peaks, B and D, at 67.8 and 78.0 minutes respectively before the Fermi trigger, where peak B has a confidence level of $4.0\sigma$ in the upper panel and is reduced to $3.5\sigma$ in the lower panel; meanwhile peak D is enhanced of $2.9\sigma$ in the upper panel to $5.1\sigma$ in the lower panel. Finally, it is also possible to see a peak (peak C) at 72.6 minutes after the Fermi trigger; it has a low confidence level of $2.9\sigma$ in the upper panel, and is enhanced to $5.1\sigma$ in the lower panel.

So far, three Fermi GBM triggers have happened within the field of view of Tupi telescopes, and the best ground level enhancements candidate to be classified as a GeV counterpart (observed at ground) of a keV
GRB observed by a space-borne GRB detector is the Fermi GRB linked with the Fermi trigger bn08112406.   

\section{conclusions}

The EGRET GRB detections \cite{hurley94} near bright BATSE bursts are compatible with the existence of a prompt gamma-ray emission in the GeV energy region. But it is unclear how high in energy this component extends, even though many models predict a fluence at GeV-TeV that is comparable at keV-MeV scales. Perhaps the keV GRB can be considered as an initial steep decay that consists in being the tail of the prompt GeV emission, or in other words, the keV GRB can be considered as an afterglow of the GeV GRB emission. In addition, the high energy (GeV) emission is fragmented by a shock front hitting an interstellar medium, leading to GeV gamma rays interacting with the infrared and cosmic microwave background radiations in interstellar space, producing different path lengths from the source to the observer and dispersion of the arrival time at the detector; in other words, they are delayed and/or anticipated with respect to the keV-MeV burst. This scenario is compatible with the two events analyzed here.

Above 100 GeV, ground-based gamma-ray astrophysics has been obtaining expressive results by means of air Cherenkov telescopes, such as Whipple \cite{punch92} and Hess \cite{braun08}, and the water Cherenkov technique to detect air showers produced by very high-energy gamma rays as they interact with the Earth's atmosphere. Of these, the most promising for the detection of GRBs is the second, such as MILAGRO \cite{abdo1,abdo2}, due to its high duty cycle, above 90\%, and its large field of view, around $2.0$ sr. However, above 100 GeV no conclusive emission has been detected for any single GRB. For instance, there were $\sim 42$ satellite-triggered GRBs within the field of view of MILAGRO and no significant emission was detected from any of these bursts. This result is an indication of a high absorption of high-energy gamma ray by the extragalactic background light or perhaps, in part, of the atmospheric absorption of air shower particles initiated by a gamma ray of $\sim 100$ GeV. In order to minimize the atmospheric absorption, the High Altitude Water Cherenkov (HAWC) project represents the next generation of water Cherenkov gamma ray detectors at extreme altitude ($>4$km asl)\cite{hawc}.

On the other hand, the shielding effect of the magnetosphere is the lowest in the SAA region and the Tupi muon telescopes have proved to be a useful tool to monitor the particle precipitation in this region \cite{augusto08a,augusto08b}. 
As the transverse magnetic field is very small in the SAA region, the lateral dispersion of the particles produced by a transient event is also smaller, and the number of particles collected by a detector is larger. In short, the sensitivity of particle telescopes is highest in the SAA region. This permits the detection of transient events of small scale, such as solar flares \cite{navia05,augusto05}, and so far the Tupi detector has proved to be a efficient tool to detect at ground, with high confidence, the GeV counterpart of keV GRBs observed in space-borne detectors,
two indicated in  \cite{augusto08c} and two analyzed here.

 We expect other simultaneous space-borne and ground GRB detections and thus to clarify several aspects of the GRB phenomenon that are still unclear. For instance, if the GeV counterpart of keV GRBs is real then
cosmological models are excluded for these bursts because the absorption the GeV to TeV gamma rays by the interstellar
background radiation.

\acknowledgments

This work is supported by the National Council of Research (CNPq) of Brazil, under Grant $479813/2004-3$ and $476498/2007-4$. We are grateful to various catalogs available on the web and to their open data policy, especially to the Fermi GBM Trigger Catalog and the GCN report.

{}
\end{document}